\documentclass[final,5p,times,twocolumn]{elsarticle}
\usepackage{graphicx}
\usepackage{epsfig}
\usepackage{stfloats}
\usepackage{fixltx2e}
\usepackage{amssymb}
\usepackage{amsmath}

\journal{Physics Letters B}

\begin{document}

\begin{frontmatter}

\title{Features of ghost-gluon and ghost-quark 
bound states related to 
BRST quartets}

\author{Natalia Alkofer and  Reinhard Alkofer}

\address{Institute of Physics, University of Graz, Universit\"atsplatz 5,
8010 Graz, Austria}

\begin{abstract}
The BRST quartet mechanism in infrared Landau gauge QCD is investigated.  Based
on the observed positivity violation for transverse gluons $A_{\mathrm {tr}}$ 
the field content of the non-perturbative BRST quartet  generated by
$A_{\mathrm {tr}}$ is derived. To identify the gluon's BRST-daughter state as
well as the Faddeev-Popov--charge conjugated second parent state, a truncated
Bethe-Salpeter equation for the gluon-ghost bound state in the  adjoint colour
representation is derived and studied. This equation is found  to be 
compatible with the so-called scaling solutions of functional approaches. 
Repeating the same construction for quarks instead of $A_{\mathrm {tr}}$ leads
in a similar way to a truncated Bethe-Salpeter equation for the quark-ghost
bound state in the fundamental representation. Within the scaling  solution  the
infrared divergence of the quark-gluon vertex is exactly the right one to make
this Bethe-Salpeter equation infrared consistent.
\end{abstract}


\end{frontmatter}

\section{Introduction}

As it is well-known by now, gauge-fixing a Yang-Mills (YM) theory leaves 
nevertheless an important invariance: The YM  action is invariant under a
transformation which may be pictured as a gauge transformation but with the
Faddeev-Popov ghost as transformation ``parameter'' 
\cite{Becchi:1974md,Tyutin:1975qk,Becchi:1996yh}. 
This BRST transformation is on the
perturbative level the corner stone in the proof of renormalizibility, for an
accurate discussion see {\it e.g.\/}  Ref.~\cite{Weinberg:1996kr}. In addition,
it can be employed to define a physical state space which in this language is
identified with the cohomology of the BRST charge operator, see Ref.\ 
\cite{Nakanishi:1990qm} and references therein. Obviously, the definition of a
positive-definite state space is intimately related to confinement: The
coloured states underlying confinement cannot be attributed to asymptotic
states in the $S$-matrix. To this end it is interesting to note that the BRST
cohomology automatically contains only colour-singlet states.

These remarks make evident that the two purposes BRST cohomology serves
 require two different kinds of BRST multiplets.\footnote{Besides the
colourless BRST singlets there are only the so-called BRST quartets, see
below.} There are on the one hand the perturbative BRST multiplets, first of
all the elementary BRST quartet \cite{Nakanishi:1990qm} which takes care of the
cancellation of longitudinal and time-like gluons as well as ghosts and
antighosts in all physical states. This is nothing else than the generalization
of the Gupta-Bleuler formalism of QED. Or, as Weinberg
\cite{Weinberg:1996kr} phrases it, the perturbative BRST quartet mechanism in
YM is an $(N_c^2-1)$-fold\footnote{$N_c^2-1$ being the dimension of the adjoint
representation.} duplication of the single cancellation mechanism in QED. For the
definition of the state space as BRST cohomology these considerations  are by
far not sufficient. On the contrary, in QCD we expect the BRST cohomology to
contain only glueballs and hadrons, {\it i.e.\/} bound states. As bound states
occur only beyond perturbation theory it becomes evident that one has to employ
non-perturbative techniques in studying the BRST cohomology in general. 

At this point it is interesting to note that recently a possibility has been
suggested \cite{vonSmekal:2008ws} to avoid the Neuberger 0/0 problem of lattice
BRST \cite{Neuberger:1986vv} in a modified lattice Landau gauge which employs 
stereographic projections.
Besides other fundamental issues elucidated in ref.\ \cite{vonSmekal:2008ws} 
the following aspect is important in the present context: It is now clarified
that the global BRST charge, necessary for the perturbative as well as for the
non-perturbative quartet 
mechanism, can be well defined.  

One interesting aspect of BRST multiplets follows from the nilpotency of the
BRST transformation, see Sect.~2 for details. Every non-singlet state can then
produce only one further state, making thus a doublet. It proves, however,
useful to form quartets such that the Faddeev-Popov charge conjugated state of
the daughter state in this doublet is used as a parent state which under BRST
generates the 2nd daughter and thus completes the quartet. In this letter we
will argue that positivity violation of transverse gluons imply that they are 
parent states. As we will show this implies that the 1st daughter has to be a
gluon-ghost bound state, and the 2nd parent accordingly a gluon-antighost bound
state. In Landau gauge which is ghost-antighost symmetric these two bound
states become degenerate and are described by the same equation. Formally, the
same remarks apply for the quartet generated by quarks. It is one of the main
results to be presented here that the structures of the corresponding
Bethe-Salpeter equations (truncated to keep the infrared leading terms) 
become very similar due to the dressing of the quark-gluon vertex.

In Sect.~2 we demonstrate briefly how to obtain the field content of the
gluon- and the quark-generated non-perturbative
BRST quartets.\footnote{The main results of this section have been known, of
course, since quite some time. However, in the way needed in the following it
is not easily accessible in the literature. A more detailed description of the
material summarized in Sect.~2 may be found in
Ref.~\cite{Alkofer:2011uh}.} In Sect.~3 we discuss the ghost-gluon
Bethe-Salpeter equation and provide arguments for a massless ghost-gluon bound
state. In Sect.~4 the ghost-quark bound state equation is given and discussed. 
 In Sect.~5 we conclude and close with an outlook.

\section{BRST quartets in Landau gauge QCD}

Throughout this letter we will only consider Landau gauge. To emphasize the
nilpotency of the BRST transformation we will work in a representation with
Nakanishi-Lautrup field $B^a$ which becomes on-shell identical to the  gauge
fixing condition, $B^a=({1}/{\xi}) \, \,\partial_\mu A_\mu^a$ where $\xi$  is
the gauge parameter of linear covariant gauges.

It is useful to picture the BRST transformation $\delta_B$ as a ``gauge
transformation'' with a constant ghost field as parameter
\begin{equation}
\begin{array}{ll}
\delta_B A^a_\mu \, =\,  \widetilde Z_3 D^{ab}_\mu c^b \, \lambda  \; , 
\quad &  
\delta_B q
\, = \,-  i g t^a \widetilde Z_1 \, c^a \, q \, \lambda \; , \\
\delta_B c^a \, = \, - \, \frac{g}{2} f^{abc} \widetilde Z_1 
 \, c^b c^c \, \, \lambda \; ,
\quad  & \delta_B\bar c^a \, = \, B^a
\, \lambda \; , 
\\
\delta_B B^a \, = \, 0,\end{array}  
\label{BRST}
\end{equation}
where $D^{ab}_\mu$ is the covariant derivative. The parameter $\lambda$ lives
in the Grassmann algebra of the ghost fields $c^a$, it carries ghost number
$N_{\mbox{\tiny FP}} = -1$. $\widetilde Z_1$ and $\widetilde Z_3$ are the
ghost-gluon-vertex and the ghost wave function renormalization constants. 
In Landau gauge one has $\widetilde Z_1 =1$.

Via the Noether theorem one may define a BRST charge operator $Q_B$ which in
turn generates a ghost number graded algebra on the fields, $\delta_B\Phi = \{
i Q_B, \Phi \}$. Defining the ghost number operator $Q_c$ one obtains
\begin{equation} 
Q_B^2 = 0 \; , \quad \left[ iQ_c , Q_B \right] = Q_B \; .
\end{equation} 
This algebra is complete in the full (indefinite metric) state space of YM 
theory, resp., QCD. The BRST cohomology is then constructed as follows: The
semi-definite  physical subspace  $\mbox{Ker}\, Q_B  $ is defined on the basis
of this algebra by those states which are annihilated by the BRST charge $Q_B$,
$Q_B |\psi \rangle =0$. Since $Q_B^2 =0 $, this subspace contains the space $
\mbox{Im}\, Q_B $ of the so-called daughter states $Q_B |\phi \rangle$ which are
images of their parent states in full  state space. 
A physical Hilbert space is then obtained as the space of
equivalence classes:
\begin{equation}
     {\mathcal{H}}(Q_B) = {\mbox{Ker}\, Q_B}/{\mbox{Im}\, Q_B}
        \; .
\end{equation}
This Hilbert space is isomorphic to the space of BRST singlets. All states are
either BRST singlets or belong to quartets, 
this exhausts all possibilities.
Note that the condition $Q_B |\psi\rangle = 0$
eliminates half of these metric partners from  all $S$-matrix elements, leaving
unpaired states of zero norm which do not contribute to any observable.

For constructing the so-called elementary quartet
\cite{Kugo:1979gm,Nakanishi:1990qm} one considers the asymptotic states related
to the time-like and longitudinal gluons as well as the ghost and the
antighost.\footnote{As the Nakanishi-Lautrup field $B^a$  relates to a linear
combination of time-like and longitudinal gluons, the so-called backward
polarization, the corresponding asymptotic ``one-gluon'' parent (daughter)
states are forwardly (backwardly) polarized, see {\it e.g.} Chapter 16 of
Ref.~\cite{Peskin:1995ev}.} Hereby one gluon polarization and the antighost
provide the parent states, the orthogonal gluon polarization and the ghost 
yield the
daughter states. In all physical states the contribution of this quartet
cancels similar to the cancellation of time-like and longitudinal photons in QED.
This quartet is strictly perturbative in the sense that it also exists in the
limit $g\to0$. From the construction of this elementary quartet one can infer
how to construct other perturbative ``multi-particle'' BRST quartets 
\cite{Nakanishi:1990qm}: Starting
from a state with negative norm (1st parent) one obtains the 1st daughter by
acting with the BRST charge operator on the 1st parent. The Faddeev-Popov charge
reflected state of the 1st daughter provides the 2nd parent. Acting on it with
the BRST charge operator provides the 2nd daughter.

At the perturbative level the transverse gluons belong to the BRST cohomology
which is, of course, in open conflict with the observed confinement of gluons.
Therefore it has been speculated already decades ago that the transverse gluons
are 
also part of a BRST quartet \cite{Kugo:1979gm} which is then in turn believed to
be an important aspect of gluon confinement. At approximately the same time it
has been observed \cite{Oehme:1980ai} that the antiscreening of gluons (which is
a very welcome property as it explains asymptotic freedom) is already at the
perturbative level in conflict with the positivity of the gluon spectral
function. By now there is no doubt any more that the transverse gluons of Landau
gauge QCD are positivity violating, see {\it e.g.} Ref.~\cite{Bowman:2007du} and
references therein. With the remarks given above this makes plain that
``one-transverse-gluon'' states are BRST parent states. Their respective
daughters, however, cannot be the elementary ``one-ghost'' states because these
are members of the elementary quartet. On inspecting eq.\ (\ref{BRST}) 
one can immediately conclude that the corresponding daughter state needs to have
the field content 
\begin{equation}
   \widetilde Z_3 f^{abc} A^c_\mu c^b \, .
\end{equation}
As for every ``one-transverse-gluon'' state there should occur exactly one
daughter state this requires the existence of a ghost-gluon bound state in the
adjoint representation. In this sense the resulting BRST quartet is strictly
non-perturbative as the formation of bound states can be described only with
non-perturbative techniques. The Faddeev-Popov charge
reflected state is then an antighost-gluon bound state. Here Landau gauge
provides an advantage as compared to general linear covariant gauges: In the
limit $\xi\to0$ the formalism becomes ghost-antighost-symmetric, and thus the
existence of a ghost-gluon bound state implies the occurrence of a degenerate
antighost-gluon bound state with same quantum numbers. Even having then the 2nd
parent, the BRST transformation (\ref{BRST}) leaves then three possibilities
for the 2nd daughter: A ghost-antighost bound state, a ghost-antighost-gluon
bound state, or a bound state of two differently polarized gluons. However, 
studying the
2nd daughter state is beyond the scope of this letter, and we will return below
to the ghost-gluon bound state.

The respective issues for quarks in the Landau gauge are much less clear. First
of all, it is not known whether quarks violate positivity. Although   for light
quarks dynamical chiral symmetry breaking (and for heavy quarks explicit chiral
symmetry breaking) determines the infrared behaviour of the quark propagator the
analytic structure of the quark propagator is highly sensitive to details in the
quark-gluon vertex,  see, {\it e.g.\/}, Ref.~\cite{Alkofer:2003jj}. The
quark-gluon vertex for light quarks is, on the other hand, also very strongly
influenced by dynamical chiral symmetry breaking
\cite{Skullerud:2003qu,Alkofer:2006gz,Alkofer:2008tt}. 
Second, an inspection of eq.~(\ref{BRST})
reveals that if a BRST quartet is generated by quarks it can only be a
non-perturbative one. Which mechanism then guarantees that the corresponding
bound states are degenerate with the quark states is completely unknown. 

Before trying to give an at least partial answer to this we will discuss the
infrared  behaviour of Landau gauge YM theory and the implications for a
possible ghost-gluon bound state and its role in the corresponding BRST quartet.

\section{Landau gauge YM theory and ghost-gluon bound state}

Over the last decade the infrared behaviour of Landau gauge YM theory has been
in the focus of many studies. Hereby it is interesting to note that in the 
deep infrared
quite general statements can be deduced by employing functional equations. On
the one hand, Dyson-Schwinger equations have been used to extend a previous
analysis of gluon and ghost propagators  
\cite{vonSmekal:1997is,Watson:2001yv,Zwanziger:2001kw,Lerche:2002ep,Fischer:2002hna} 
to all Yang-Mills vertex functions \cite{Alkofer:2004it,Huber:2007kc}. Employing 
in addition the Function Renormalization Group Equations, and requiring that these
two, seemingly different, towers of equations have to provide identical Green's
functions, allows a powerful restriction on the type of the solution: 
There is one
unique scaling solution with power laws for the Green's functions
\cite{Fischer:2006vf,Fischer:2009tn} and a one-parameter family of solutions,
the so-called decoupling solutions. These infrared trivial
solutions possess as an endpoint exactly the scaling solution which is
characterized by infrared power laws.  Numerical
solutions of the decoupling type (there called ``massive solution'')
have been published in \cite{Aguilar:2008xm,Boucaud:2008ky} and
references therein. A recent detailed description and comparison of
these two types of solutions has been given in Ref.\
\cite{Fischer:2008uz}, see also Refs.
~\cite{Alkofer:2008jy,Huber:2009wh,Szczepaniak:2001rg,Epple:2007ut}.

Although almost all lattice calculations of the gluon propagator favor a
decoupling solution  lattice studies at strong coupling
\cite{Sternbeck:2008mv,Maas:2009ph,Cucchieri:2009zt} reveal the existence of a
regime where the scaling relation between the gluon and the ghost propagators is
fulfilled, and the corresponding infrared exponent $\kappa$ is very close to the
value determined in truncated continuum studies with $\kappa=0.595$. A 
potential explanation of these conflicting results 
has been discussed recently \cite{Maas:2009se}: The
infrared behaviour of the Green's functions may depend on the
non-perturbative completion of the gauge.

In the present context it is important to note that the scaling solution
respects BRST symmetry whereas every decoupling solution breaks it
\cite{Fischer:2008uz}, although very likely only softly. In a strict sense our
analysis as presented in the following will be only valid if the scaling
solution is a correct one. However, the situation is not as drastic as it may
seem. First, if the conjecture of Ref.~\cite{Maas:2009se} is correct it is
sufficient that only one non-perturbative completion of Landau gauge with
scaling solution exists to make our analysis well-founded. Second, even if only
decoupling type of solutions will turn out to be correct an extended BRST-like
nilpotent symmetry is likely to take the role of the BRST symmetry 
\cite{Sorella:2009vt}, or the soft  BRST symmetry breaking can be treated as 
spontaneous symmetry breaking \cite{Zwanziger:2010iz}, see also
Ref.~\cite{Sorella:2011tu} and references therein. Note also
that all our arguments about infrared dominance of diagrams will likely
stay  correct: 
The  numerical value of a diagram which is infrared leading in the scaling
solution will be enhanced ({\it i.e.\/} numerically large)
in a decoupling solution if the latter is not  
far from the endpoint given by the scaling solution. 

As a basis for the following discussion
we will summarise the infrared behaviour of all one-particle
irreducible Green's functions in the scaling solution in the simplified case 
with only one external spacelike scale $p^2\to 0$. For a function with 
$n$ external ghost and antighost as well as $m$ gluon legs one has:
\begin{equation}
\Gamma^{n,m}(p^2) \sim (p^2)^{({n-m})\kappa} .
\end{equation}
This solution fulfills all functional equations and all
Slavnov-Taylor identities. It verifies the hypothesis of infrared
ghost dominance \cite{Zwanziger:2003cf} and leads to an infrared diverging 
ghost propagator as well as infrared diverging three- and 
four-gluon vertex functions.

As  already stated gluons violate positivity
\cite{Bowman:2007du,Alkofer:2003jj}. For the scaling solution this can be
immediately deduced from the fact that for this solution the gluon propagator
vanishes at zero virtuality, $p^2=0$. A further important property of the
scaling solution is the infrared trivial behaviour of the ghost-gluon vertex
which is in agreement with general arguments \cite{Taylor:1971ff,Lerche:2002ep}.

To search for the anticipated ghost-gluon bound state we want to truncate the
ghost-ghost-gluon-gluon scattering kernel to the infrared leading term. To this
end we employ first the MATHEMATICA package DoDSE, resp., DoFun 
\cite{Alkofer:2008nt,Huber:2010ne} to
derive the diagrammatic expressions for the one-particle irreducible
ghost-ghost-gluon-gluon scattering kernel.  The diagram-by-diagram infrared
power counting\footnote{Some more details of this analysis are given in ref.\
\cite{Alkofer:2011uh}.} verifies that in the scaling solution the infrared
exponent is $-\kappa$. It also provides the infrared leading terms.

\begin{figure*}[t]
\begin{center}
\includegraphics[scale=0.95]{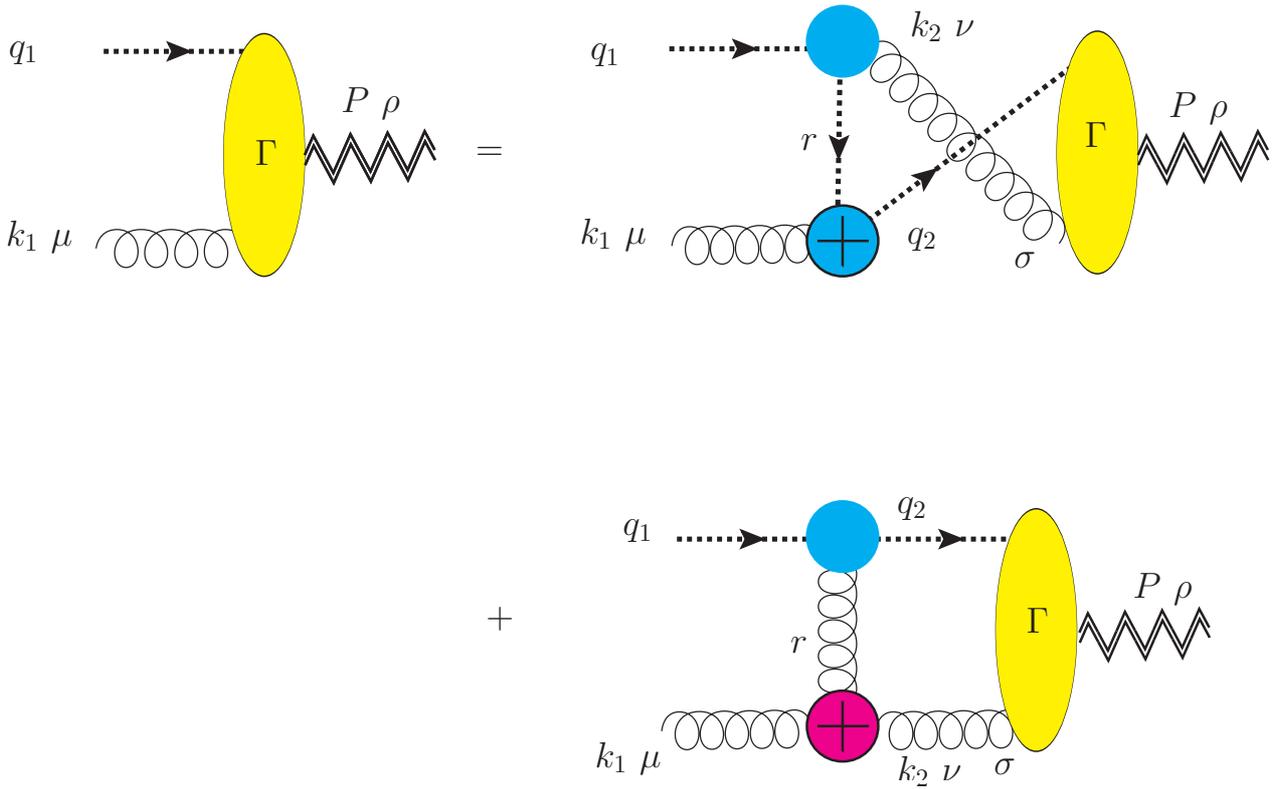}
\end{center}
\caption{Graphical representation of the gluon-ghost Bethe-Salpeter equation. A
cross denotes a dressed vertex.
\label{GhGl}}
\end{figure*}
As two different kind of fields are
involved there exists two distinct possibilities for the Dyson-Schwinger
equation according to which leg one puts the bare vertex. Placing the bare
vertex on a gluon line gives 56 diagrams (59 if quarks are included) with six
(seven) of them containing  5-point or 6-point functions. Eleven of them are
infrared leading. Assigning the bare vertex to a ghost leg leads to a 
Dyson-Schwinger equation with 13 diagrams on the r.h.s, one containing a 5-point
function, and all of them infrared leading. As we have to truncate the system
such to neglect $n\ge 5$-point functions it is obvious that the second choice
minimizes the truncation error.\footnote{Note that either neglecting 
or approximating the one-particle irreducible  5-point function is necessary to
obtain a solvable Bethe-Salpeter equation. However, as the diagram containing
the 5-point function possesses the same infrared exponent as the other diagrams
the resulting equation stays qualitatively correct, and it is reasonable to
assume that it is quantitatively satisfactory. Please note also that the
truncation is not based on the number of diagrams but based on the argument to
keep as many infrared leading diagrams as technically possible. } 
This is substantiated by the fact that the first
choice would provide eventually a Bethe-Salpeter equation with one less term as
if we will employ the second choice, see below.

The requirements for the diagrams on the r.h.s to be kept are: It should contain
the one-particle irreducible ghost-ghost-gluon-gluon four-point function and no
$n\ge 5$-point function, it should be infrared leading, and it should contain
the interaction in the ghost-gluon channel. This leaves two diagrams:
(i) one with two ghost and one gluon propagator on
internal lines. This is effectively a ghost exchange. (NB: A more
precise description is that the leading tree-level diagram describes the
splitting of the incoming gluon into a ghost-antighost pair and a fusion of the
incoming ghost with the exchanged (anti-)ghost to a gluon, {\it cf.\/} the upper
r.h.s of Fig.~\ref{GhGl}.) 
(ii) one with two gluon and one ghost propagators on
internal lines. This is a gluon exchange. Note that this diagram is infrared
leading because in the scaling solution the fully dressed
three-gluon vertex is infrared divergent with an exponent
$-3\kappa$. If we had chosen to put the bare vertex on the gluon leg ({\it
cf.\/} the first choice above) this diagram would be infrared suppressed.
We would also like to remark that from the 14 tensor structures of the 
three-gluon vertex at least ten 
are infrared divergent with the same exponent \cite{Alkofer:2008dt}.

Assuming the existence of a bound state as well as employing the
usual decomposition of the (ghost-ghost-gluon-gluon) 
four-point function into Bethe-Salpeter amplitudes and performing the expansion
around the pole (see {\it e.g.\/} Sect.~6.1 of Ref.~\cite{Alkofer:2000wg}) one
arrives at the Bethe-Salpeter equation depicted in Fig.~\ref{GhGl}.
Using the propagator
parameterizations of {\it e.g.\/} Ref.\ \cite{Alkofer:2003jj}, the
ghost-gluon vertex of Ref.\ \cite{Schleifenbaum:2004id}, and the three-gluon
vertex of Ref.\ \cite{Alkofer:2008dt} one can derive a self-consistent 
equation for the corresponding Bethe-Salpeter amplitude containing otherwise 
only known quantities. This leads, mostly due to the many tensor terms of the 
three-gluon vertex, to very lengthy expressions which will be given elsewhere. 
For the illustrational purposes of this letter we will 
neglect the gluon-exchange term and keep only the ghost exchange. 

As stated above in Landau gauge the ghost-gluon vertex stays even in the scaling
solution infrared trivial. Thus  it is sufficient
to consider only bare ghost-gluon vertices, {\it i.e.\/} it is sufficient to
consider the ladder approximation to the first term of the
above  Bethe-Salpeter equation.
(We also use $\widetilde Z_1=1$ \cite{Taylor:1971ff}.)
The explicit expression for the kernel, assuming as gauge group SU($N_c$) and
working in Euclidean momentum space, reads then
\begin{eqnarray}
\delta^{ab} H_{\mu \nu} (k_1,q_1;k_2,q_2) &=& f^{acd} f^{bef} ig f^{edg} 
(q_1-k_2)_\nu \nonumber \\ 
&& \quad D_G(q_1-k_2) ig f^{cfg} q_{2\mu} \\
&=& - g^2 N_c^2 \delta^{ab} r_\nu q_{2\mu} D_G(r),
\nonumber
\end{eqnarray}
where we have already taken care of the re-projection onto the adjoint
representation and have used the momentum assignment as in the r.h.s of
Fig.~\ref{GhGl}. Hereby $D_G(q_1-k_2)=D_G(r)$ is the ghost propagator.

With this kernel and denoting the gluon propagator as $D_{\mu \nu } (k)$
we arrive at the gluon-ghost Bethe-Salpeter equation for  bound state with
four-momentum $P$
\begin{eqnarray}
\Gamma_{\mu \rho } (k_1,q_1;P) &=& - \widetilde Z_3^2 g^2 N_c^2 
\int \frac {d^4r}{(2\pi)^4}  r_\nu q_{2\mu} D_G(r) \label{GhGlBS}\\
&& \qquad \quad D_{ \nu \sigma} (k_2)
D_G(q_2) \Gamma_{\sigma \rho } (k_2,q_2;P). \nonumber 
\end{eqnarray}
As a side remark we want to mention that to arrive at the equation for the 
antighost-gluon bound state we only need to invert all ghost momenta and obtain
therefore:
\begin{eqnarray}
\widetilde \Gamma_{\mu \rho } (k_1,q_1;P) &=& - \widetilde Z_3^2 g^2 N_c^2 
\int \frac {d^4r}{(2\pi)^4} q_{1\nu} r_\mu  D_G(r) \label{aGhGlBS}\\
&& \qquad \quad D_{ \nu \sigma} (k_2)
D_G(q_2) \widetilde \Gamma_{\sigma \rho } (k_2,q_2;P). \nonumber 
\end{eqnarray}

As argued above we are looking for a massless bound state. This allows to
specialise to the soft limit $P\to 0$. Employing the transversality of the 
Landau gauge gluon propagator, 
\begin{equation}
 D_{ \nu \sigma} (k) = \left(\delta _{ \nu \sigma} - \frac{k_\nu k_\sigma}{k^2}
 \right) \frac{Z(k^2)}{k^2}
\end{equation}
it is straightforward to show that $\Gamma_{\mu \rho } (k,-k;0)$ (resp.,
$\widetilde \Gamma_{\mu \rho }  (k,k;0$) is transverse, too:
\begin{equation}
\Gamma_{\mu \rho } (k,-k;0) =: \left(\delta _{ \mu \rho } - 
\frac{k_\mu k_\rho}{k^2}  \right) {F(k^2)}.
\end{equation}
Defining the ghost renormalization function $G(q^2)=q^2D_G(q^2)$ one obtains
then from eq.\ (\ref{GhGlBS})
\begin{eqnarray}
F(k_1^2) &=&  \widetilde Z_3^2 g^2 N_c^2 
\int \frac {d^4k_2}{(2\pi)^4} \frac{G((k_1+k_2)^2)}{(k_1+k_2)^2}
\frac{G(k_2^2)}{k_2^2} \frac{Z(k_2^2)}{k_2^2}
 \nonumber \\
&& \qquad   \frac 1 3 (k_1\cdot k_2) 
\left( 1 - \frac {(k_1\cdot k_2)^2}{k_1^2 k_2^2}\right) F(k_2^2).
\label{GhGlBSforF}
\end{eqnarray}
It is an easy exercise to verify that the Bethe-Salpeter equation 
(\ref{aGhGlBS}) provides exactly the same equation as (\ref{GhGlBSforF}) and
therefore the degeneracy of the massless antighost-gluon bound state if the
massless gluon-ghost bound state exists. 

The numerical solution of eq.\ (\ref{GhGlBSforF}) is beyond the scope of the
present letter, it will be presented elsewhere. Here we only want to  mention
that, first, $F(k^2)$ has the peculiar feature $F(0)=0$, and, second, due to
truncation errors we do not expect the equation to be exactly fulfilled. As,
however, we have kept the infrared leading term it is important to test whether
there is no contradiction on the analytical level.

For the scaling solution we have the power laws 
\begin{equation}
Z(p^2) \sim (p^2)^{2\kappa}, \qquad G(p^2) \sim (p^2)^{-\kappa}, 
\label{ScalProp}
\end{equation}
which implies that the kernel of eq.\ (\ref{GhGlBSforF})
possesses a vanishing anomalous infrared exponent and is independent of the
infrared exponent $\kappa$. We therefore get the expected result that the
Bethe-Salpeter equation for a massless (anti-)ghost-gluon bound state is
consistent within the infrared analysis of the scaling solution of functional
equations. Furthermore, as the combination 
$\alpha^{\mathrm gh}(p^2) =\frac{g^2}{4\pi}G^2(p^2) Z(p^2)$
has the form of a running coupling with an infrared fixed point 
$\alpha^{\mathrm gh}(0) = 8.92/N_c$ (see {\it e.g.\/} Sect.~2.3 of 
Ref.~\cite{Fischer:2006ub}) it is plausible that the kernel of eq.\
(\ref{GhGlBSforF}) is strong enough to lead to a bound state. 
All these findings 
corroborates the validity of the employed method, and therefore we
will investigate the quark-gluon bound state with the same method.

\section{Quark-gluon bound state}

\begin{figure*}[t]
\begin{center}
\includegraphics[scale=0.95]{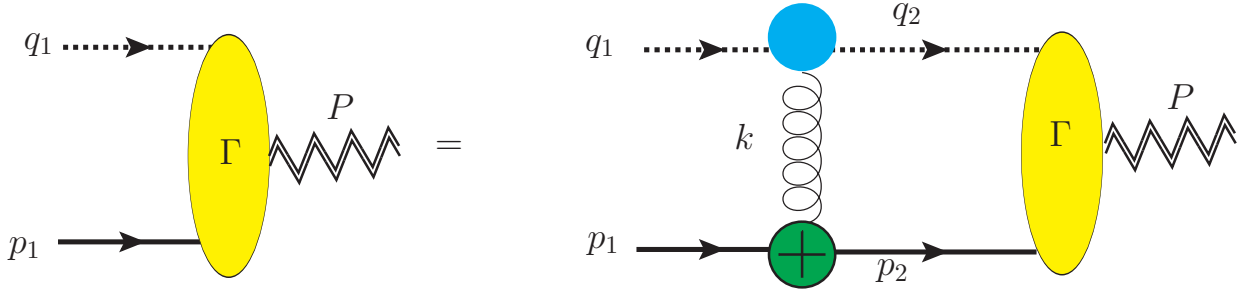}
\end{center}
\caption{Graphical representation of the quark-ghost Bethe-Salpeter equation.
Note that the quark-gluon vertex is fully dressed.
\label{GhQu}}
\end{figure*}
The scaling solution for the YM Green's functions leads to dynamical chiral
symmetry breaking in the quark sector \cite{Alkofer:2008tt}. The quark
propagator which is of the general form
\begin{equation}
S(p)
= \frac{i  p\hspace{-.5em}/\hspace{.15em} + M(p^2)}{p^2 + M^2(p^2)}Z_f(p^2)
\label{QuarkProp}
\end{equation}
is then infrared finite, $S(0)= Z_f(0)/M(0)$. The twelve possible Dirac
tensor structures of the quark-gluon vertex are then all infrared divergent with
an infrared exponent $-\kappa -1/2$. As a side remark we want to add that the
same infrared divergence results for vanishing gluon momentum, and that this
leads to $1/k^4$ behaviour of the kernel in the four-quark function, $k$ being
the momentum exchange. This is indicative of a linearly rising  potential
between static quarks, and thus quark confinement. 

Furthermore, the Slavnov-Taylor identities require that the
ghost-ghost-quark-quark scattering kernel is infrared trivial,
see Sect.~3.9 in Ref.~\cite{Alkofer:2008tt}. This information is absolutely 
necessary for the following analysis. 

As in the previous subsection we have two choices for the Dyson-Schwinger
equation according to which leg one puts the bare vertex. 
Both choices lead to seven diagrams on the r.h.s of the equation, in both
choices there appears one diagram with a 5-point function
(ghost-ghost-quark-quark-gluon).
However, choosing the quark leg to carry the bare vertex only one diagram is
infrared leading, namely exactly the one with the 5-point function. Thus
truncating the equation to $n\le4$-point functions would lead to a 
contradiction. Choosing a ghost leg to place the bare vertex there are four
infrared leading diagrams: One with the  5-point function, and three according
to the three possible ($s$,$t$,$u$) interaction channels. Therefore, employing
the same truncation requirements and the same derivation of the Bethe-Salpeter
equation as in the previous subsection one arrives at the equation depicted in
Fig.~\ref{GhQu}. This equation is in full agreement with the infrared analysis
of the scaling solution, {\it i.e.\/} it is a valid bound state equation, and 
in its kernel the infrared exponent $\kappa$ cancels. 

Employing all twelve tensor structures of the quark-gluon vertex does generate
a very lengthy equation. Again we will restrict for illustrational purposes to
the perturbatively leading tensor component,
\begin{equation}
\Gamma_\mu^a (p_1,p_2;k) \to -i g \frac{\lambda^a}2 \gamma_\mu V(p_1,p_2;k)
\end{equation} 
where $V$ diverges like $(k^2)^{-\kappa -1/2}$ for $k^2 \to 0$.
Using the result 
\begin{equation}
\frac{\lambda^a}2 \frac{\lambda^c}2 \frac{\lambda^b}2 f^{abc} = - \frac i 4
(N_c^2-1) 
\end{equation} 
valid for SU($N_c$) one can project the bound state onto the fundamental colour
representation. The Bethe-Salpeter equation for the ghost-quark bound state
amplitude with total momentum $P$ reads then:
\begin{eqnarray}
\Gamma (p_1,q_1;P) &=& \frac i 4 (N_c^2-1) \int \frac {d^4k}{(2\pi)^4} 
q_{2\mu}  D_{ \mu \nu} (k) \gamma_\nu \label{GhQuBS}
\\
&& \quad  V(p_1,p_2;k) S(p_2) D_G(q_2) 
\Gamma (p_2,q_2;P) \nonumber 
\end{eqnarray}
with $p_2=p_1+k$ and $q_2=q_1-k$.
At first sight it may look like that the Bethe-Salpeter amplitude of this
scalar-fermion bound state has four independent amplitudes. This is, however, 
not correct. Should the bound state occur at a non-vanishing 
$P^2=M_B^2<0$ (to avoid
saying the ``ghost-quark bound state is massive'') there will exist two
degenerate solutions with respectively positive and negative energy. Then it is
sufficient to study only the positive-energy solution by explicitly putting a 
positive-energy  projector $\Lambda^+$ into the tensor
decomposition\footnote{This is very similar to the bound state equation of a
scalar diquark and a quark for a nucleon, see {\it e.g.\/} Sect.~7.3 of
Ref.\ \cite{Alkofer:2000wg}.} 
\begin{equation}
\Gamma (p,q;P) = \left( H_1^\prime + \frac i {M_B} 
( p\hspace{-.5em}/\hspace{.15em} -  q\hspace{-.5em}/\hspace{.15em})
H_2^\prime \right) \Lambda^+ , \quad \Lambda^+ = 
\frac 1 2 (1+ 
\frac { P\hspace{-.5em}/\hspace{.15em}}{iM_B}).
\label{FullAnsatz}
\end{equation}
 The Dirac algebra to generate the set of coupled equations for $ H_1^\prime$
  and
$H_2^\prime $ is lengthy but  straightforward. However, as the equations
themselves are somewhat vast we refrain from displaying them here. 

Again we want to simplify for illustrational purposes and simply set $\Gamma
(p,q;P) = H(p,q;P)$, {\it i.e.\/} to a scalar function. In the case of the
dominance of the ``upper Dirac component'' this should be a reasonable
approximation. To further demonstrate the properties of the kernel we make the
simplest possible choice for $V(p_1,p_2;k)$ agreeing with its infrared limit:
$V(p_1,p_2;k) \to G(k^2)/\sqrt{k^2}$.
The corresponding equation for the simplified amplitude $H(p,q;P)$ reads then:
\begin{eqnarray}
H(p_1,q_1;P)  &=& g^2 \frac {N_c^2 - 1} 4 
\int \frac {d^4k_2}{(2\pi)^4} \frac{Z(k^2)}{k^2} \frac{G(k^2)}{\sqrt{k^2} }
\frac{G(q_2^2)}{q_2^2} \nonumber \\
&&  
  \frac{Z_f(p_2^2)}{p_2^2 + M^2(p_2^2)} \left( q_1\cdot p_1
 - \frac{q_1\cdot p_1 p_1\cdot p_2}{k^2} \right) \nonumber  \\ && 
 \qquad  \qquad \qquad 
H(p_2,q_2;P). \label{GhQuBSforF} 
\end{eqnarray}
The structural similarities to eq.\ (\ref{GhGlBSforF}) are striking. Again it is
likely that the kernel is strong enough to lead to a bound state. Providing
further evidence that these bound states are degenerate with quark states will,
however, require the study of the system with the full ansatz
(\ref{FullAnsatz}).

At the end of this section we want to add a cautionary remark:
Although the statements given in this section do not hold only in the quenched
limit of QCD but also for QCD with dynamical fermions, the resulting equations
cannot be used directly. Note that {\it e.g.\/} in the step from eq.\
(\ref{GhQuBS}) to eq.\ (\ref{GhQuBSforF}) a result for the quark-gluon vertex 
\cite{Alkofer:2008tt} has been used which is only known in the quenched limit.

\section{Conclusions and outlook}

The observed positivity violation of the Landau gauge gluon propagator imply
that, if BRST is an unbroken symmetry, the transverse gluons generate a
non-perturbative BRST quartet. The other three members are bound states. For
two of these we have derived a (truncated) bound state equation by analysing
the one-particle-irreducible ghost-ghost-gluon-gluon four-point function.  The
corresponding Bethe-Salpeter equation containing in its kernel infrared
divergent Green's function  is infrared consistent, a result which is far from
being trivial. In addition, we have provided arguments that the kernel is
strong enough such that a bound state is formed. All these findings corroborate
the existence of the non-perturbative BRST quartet with transverse gluons as
first parent states. On the other hand, one might also argue that the employed
method is consistent and can therefore be extended to study the analogue
problem for quarks.

With the same method we derive then the bound state equation for a quark-ghost
bound state truncated to the infrared leading term. We note that this equation
is within the scaling solution  only then infrared consistent if the recently
found infrared divergence of the fully dressed quark-gluon vertex  is taken
into account. We take this as further evidence that the infrared behaviour of
the quark-gluon vertex  is intimately related to the issue of quark
confinement. Nevertheless, further studies of the quark-gluon vertex are highly
desirable  to generate progress on the question  whether in Landau gauge QCD
there is a non-perturbative  BRST quartet generated by quarks.

The numerical solution of the gluon-ghost bound state equation 
(\ref{GhGlBSforF}) can be done with standard
techniques, it will be published elsewhere. In order to solve numerically 
the quark-ghost bound state equation 
(\ref{GhQuBS}) and its analogue for the quark-antighost bound state the full
Dirac tensor ansatz (\ref{FullAnsatz}) has to be plugged in, and the
corresponding Dirac algebra has to be performed. However, although the resulting
set of equations are lengthy they can be solved with methods which are by now
standard. 

Here we did not touch on the question of the respective second BRST daughters.
Following the same arguments as employed here the corresponding tasks should be
doable. However, before starting an investigation in this direction it might be
worthwhile to repeat the presented investigation with Functional Renormalization
Group Equations to minimize further or even eliminate truncation errors.

\bigskip

\section*{Acknowledgments}

We are grateful to Joannis Papavassiliou to convince us to publish this
material. We thank Dan Zwanziger for his comments on the notes of one of us (RA)
although they have been made many years ago. We acknowledge helpful discussions
with Markus  Huber (and besides the physics discussion especially for 
the hints for using DoDSE), 
Jan Pawlowski and Lorenz von Smekal. We thank 
Markus  Huber, Silvio  Sorella  and Dan Zwanziger for a critical reading of the 
manuscript.


\begin{thebibliography}{99}

%
\bibitem{Becchi:1974md}
  C.~Becchi, A.~Rouet and R.~Stora,
  Commun.\ Math.\ Phys.\  {\bf 42} (1975) 127.

%
\bibitem{Tyutin:1975qk}
  I.~V.~Tyutin,
  LEBEDEV preprint 75-39
  [arXiv:0812.0580 [hep-th]].

%
\bibitem{Becchi:1996yh}
  C.~Becchi, Lectures given at the ETH Z\"urich, May 22 - 24, 1996
  [arXiv:hep-th/9607181].

%
\bibitem{Weinberg:1996kr}
  S.~Weinberg,
  ``The quantum theory of fields. Vol. 2: Modern applications,''
{\it  Cambridge, UK: Univ. Pr. (1996) 489 p.}


%
\bibitem{Nakanishi:1990qm}
  N.~Nakanishi and I.~Ojima,
  World Sci.\ Lect.\ Notes Phys.\  {\bf 27} (1990) 1.

%
\bibitem{vonSmekal:2008ws}
  L.~von Smekal,
  Plenary talk given at the 13th International Conference on Selected 
  Problems of Modern Theoretical Physics (SPMTP08),     
  Dubna, Russia, 23 -27 June, 2008
  [arXiv:0812.0654 [hep-th]];
%
\newline
  L.~von Smekal, A.~Jorkowski, D.~Mehta, A.~Sternbeck,
  PoS {\bf CONFINEMENT8 } (2008)  048.
  [arXiv:0812.2992 [hep-th]];
\newline  L.~von Smekal,
  Invited key lecture given at the 49th Winter School in Schladming,  
  Styria, Austria, 
  26 February - 5 March 2011, to appear in the proceedings.

%
\bibitem{Neuberger:1986vv}
  H.~Neuberger,
  Phys.\ Lett.\  B {\bf 175} (1986) 69;
%
 {\it ibid.} {\bf 183} (1987) 337.

\bibitem{Alkofer:2011uh}
  N.~Alkofer and R.~Alkofer,
  PoS {\bf FACESQCD} (2011) 043
  [arXiv:1102. 3119 [hep-th]].



\bibitem{Kugo:1979gm}
 T.~Kugo and I.~Ojima, Prog.~Theor.~Phys.~Suppl. {\bf 66}, 1 (1979).


\bibitem{Peskin:1995ev}
  M.~E.~Peskin and D.~V.~Schroeder,
  ``An Introduction To Quantum Field Theory,''
{\it  Reading, USA: Addison-Wesley (1995) 842 p.}


\bibitem{Oehme:1980ai}
R.~Oehme and W.~Zimmermann,
Phys.\ Rev.\ D {\bf  21},  471 (1980).


\bibitem{Bowman:2007du}
  P.~O.~Bowman {\it et al.},
  Phys.\ Rev.\ D {\bf 76} (2007) 094505
  [arXiv:hep-lat/0703022].

\bibitem{Alkofer:2003jj}
  R.~Alkofer, W.~Detmold, C.~S.~Fischer, P.~Maris,
  Phys.\ Rev.\  D{\bf 70} (2004) 014014
  [hep-ph/0309077]; 
  \newline
  Nucl.\ Phys.\ Proc.\ Suppl.\  {\bf 141} (2005) 122
  [hep-ph/0309078].

\bibitem{Skullerud:2003qu}
  J.~I.~Skullerud {\it et al.},
  JHEP {\bf 0304} (2003) 047
  [arXiv:hep-ph/0303176].


\bibitem{Alkofer:2006gz}
  R.~Alkofer, C.~S.~Fischer and F.~J.~Llanes-Estrada,
  Mod.\ Phys.\ Lett.\   {\bf A23} (2008) 1105
  [arXiv:hep-ph/ 0607293].


\bibitem{Alkofer:2008tt}
  R.~Alkofer, C.~S.~Fischer, F.~J.~Llanes-Estrada, and K.~Schwenzer,
  Annals Phys.\  {\bf 324} (2009) 106
  [arXiv: 0804.3042 [hep-ph]].


\bibitem{vonSmekal:1997is}
  L.~von Smekal, R.~Alkofer and A.~Hauck,
  Phys.\ Rev.\ Lett.\  {\bf 79} (1997) 3591
  [arXiv:hep-ph/9705242].

\bibitem{Watson:2001yv}
  P.~Watson and R.~Alkofer,
  Phys.\ Rev.\ Lett.\  {\bf 86} (2001) 5239
  [arXiv:hep-ph/0102332].

%
\bibitem{Zwanziger:2001kw}
  D.~Zwanziger,
  Phys.\ Rev.\  D {\bf 65} (2002) 094039
  [arXiv:hep-th/0109224].


\bibitem{Lerche:2002ep}
C.~Lerche and L.~von Smekal,
Phys.\ Rev.\ D {\bf 65}  (2002) 125006
 [arXiv:hep-ph/0202194].

%
\bibitem{Fischer:2002hna}
  C.~S.~Fischer and R.~Alkofer,
  Phys.\ Lett.\  B {\bf 536} (2002) 177
  [arXiv:hep-ph/0202202].

\bibitem{Alkofer:2004it}
  R.~Alkofer, C.~S.~Fischer and F.~J.~Llanes-Estrada,
  Phys.\ Lett.\ B {\bf 611} (2005)  279
  [arXiv:hep-th/0412330].

\bibitem{Huber:2007kc}
  M.~Q.~Huber, R.~Alkofer, C.~S.~Fischer and K.~Schwenzer,
  Phys.\ Lett.\  B {\bf 659} (2008) 434
 [arXiv:0705.3809 [hep-ph]].

\bibitem{Fischer:2006vf}
  C.~S.~Fischer and J.~M.~Pawlowski,
  Phys.\ Rev.\  D {\bf 75} (2007) 025012
  [arXiv:hep-th/0609009].

%
\bibitem{Fischer:2009tn}
  C.~S.~Fischer and J.~M.~Pawlowski,
  Phys.\ Rev.\  D {\bf 80} (2009) 025023
  [arXiv:0903.2193 [hep-th]].


\bibitem{Aguilar:2008xm}
  A.~C.~Aguilar, D.~Binosi and J.~Papavassiliou,
  Phys.\ Rev.\  D {\bf 78} (2008) 025010
  [arXiv:0802.1870 [hep-ph]].

\bibitem{Boucaud:2008ky}
 P.~Boucaud {\it et al.},
  JHEP {\bf 0806} (2008) 099
  [arXiv: 0803.2161 [hep-ph]].

\bibitem{Fischer:2008uz}
  C.~S.~Fischer, A.~Maas and J.~M.~Pawlowski,
  Annals Phys.\  {\bf 324} (2009) 2408
  [arXiv:0810.1987 [hep-ph]].


\bibitem{Alkofer:2008jy}
  R.~Alkofer, M.~Q.~Huber and K.~Schwenzer,
  Phys.\ Rev.\  D {\bf 81} (2010) 105010
  [arXiv:0801.2762 [hep-th]].

\bibitem{Huber:2009wh}
  M.~Q.~Huber, K.~Schwenzer and R.~Alkofer,
  Eur.\ Phys.\ J.\  {\bf C68 } (2010)  581
  [arXiv:0904.1873 [hep-th]].

\bibitem{Szczepaniak:2001rg}
  A.~P.~Szczepaniak and E.~S.~Swanson,
  Phys.\ Rev.\  D {\bf 65} (2002) 025012
  [arXiv:hep-ph/0107078].

\bibitem{Epple:2007ut}
  D.~Epple, H.~Reinhardt, W.~Schleifenbaum and A.~P.~Szczepaniak,
  Phys.\ Rev.\  D {\bf 77} (2008) 085007
  [arXiv:0712.3694 [hep-th]].

\bibitem{Sternbeck:2008mv}
  A.~Sternbeck and L.~von Smekal,
  Eur.\ Phys.\ J.\  C {\bf 68} (2010) 487
  [arXiv:0811.4300 [hep-lat]].

\bibitem{Maas:2009ph}
  A.~Maas {\it et al.},
  Eur.\ Phys.\ J.\  C {\bf 68} (2010) 183
  [arXiv: 0912.4203 [hep-lat]].

\bibitem{Cucchieri:2009zt}
  A.~Cucchieri and T.~Mendes,
  Phys.\ Rev.\  D {\bf 81} (2010) 016005
  [arXiv: 0904.4033 [hep-lat]].

\bibitem{Maas:2009se}
  A.~Maas,
  Phys.\ Lett.\  B {\bf 689} (2010) 107
  [arXiv:0907. 5185 [hep-lat]].

%
\bibitem{Zwanziger:2003cf}
  D.~Zwanziger,
  Phys.\ Rev.\  D {\bf 69} (2004) 016002
  [arXiv: hep-ph/0303028].

\bibitem{Taylor:1971ff}
J.~C. Taylor, {Nucl. Phys.} {\bf B33} (1971) 436.


 \bibitem{Alkofer:2008nt}
  R.~Alkofer, M.~Q.~Huber and K.~Schwenzer,
  Comput.\ Phys.\ Commun.\  {\bf 180} (2009) 965
  [arXiv:0808.2939 [hep-th]];
\newline
  M.~Q.~Huber and J.~Braun,
  arXiv:1102.5307 [hep-th].

\bibitem{Huber:2010ne}
  M.~Q.~Huber, Ph.D. Thesis, University Graz, 2010
  [arXiv:1005.1775 [hep-th]].


%
\bibitem{Sorella:2009vt}
  S.~P.~Sorella,
  Phys.\ Rev.\  D {\bf 80} (2009) 025013
  [arXiv:0905.1010 [hep-th]].

%
\bibitem{Zwanziger:2010iz}
  D.~Zwanziger,
  Phys.\ Rev.\  D {\bf 81} (2010) 125027
  [arXiv:1003.1080 [hep-ph]].

%
\bibitem{Sorella:2011tu}
  S.~P.~Sorella, D.~Dudal, M.~S.~Guimaraes and N.~Vandersickel,
  PoS {\bf FACESQCD} (2011) 022
  [arXiv:1102.0574 [hep-th]].

%
\bibitem{Alkofer:2008dt}
  R.~Alkofer, M.~Q.~Huber and K.~Schwenzer,
  Eur.\ Phys.\ J.\  C {\bf 62} (2009) 761
  [arXiv:0812.4045 [hep-ph]];
  M.~Q.~Huber, R.~Alkofer and K.~Schwenzer,
  PoS {\bf CONFINEMENT8 } (2008) 174
  [arXiv:0812.4451 [hep-ph]].

 \bibitem{Alkofer:2000wg}
  R.~Alkofer and L.~von Smekal,
  Phys.\ Rept.\  {\bf 353} (2001) 281
  [arXiv:hep-ph/0007355].

%
\bibitem{Schleifenbaum:2004id}
  W.~Schleifenbaum, A.~Maas, J.~Wambach and R.~Alkofer,
  Phys.\ Rev.\  D {\bf 72} (2005) 014017
  [arXiv:hep-ph/0411052].
 
\bibitem{Fischer:2006ub}
  C.~S. Fischer, J. Phys. G: Nucl. Part. Phys. {\bf 32} (2006) R253
  [arXiv:hep-ph/0605173].


\end{thebibliography}
\end{document}